\documentclass[12pt]{article}
\usepackage{amsmath}
\usepackage{multicol}
\usepackage[dvips]{graphicx}
\usepackage{times}
\usepackage{graphicx}

\newcommand{\f}{\frac}

\newcommand{\bb}{\bibitem}
\newcommand{\BF}{\begin{figure}\begin{center}}
\newcommand{\EF}{\end{center}\end{figure}}
\newcommand{\BE}{\begin{equation}}
\newcommand{\EE}{\end{equation}}
\newcommand{\BEA}{\begin{eqnarray}}
\newcommand{\EEA}{\end{eqnarray}}

\newcommand{\IG}{\includegraphics}

\newcommand{\ms}{M_{\odot}}
\begin{document}

\title{Direct Detection of Intermediate Mass Compact Objects via
Submillilensing}
\author{Kaiki Taro Inoue and Masashi Chiba}
\date{}

\maketitle
\begin{abstract}
A galaxy-sized halo may contain 
a large number of 
intermediate mass $10^{2-4}\ms$ compact objects (IMCOs), 
which can be intermediate mass black holes (IMBHs) or
the CDM subhalos. 
We propose to directly detect the IMBHs
by observing multiply imaged QSO-galaxy
lens systems with a high angular resolution ($\sim 0.03$mas), 
which would be achieved by the next-VLBI space missions.
The silhouette of the IMBHs would appear as an either 
monopole-like or dipole-like variation at the scale of the 
Einstein radius against the QSO jets. 
As a byproduct, we can also directly
detect the $10^{4-5}\ms$ CDM subhalos.
From a measurement of the local distortion in the 
surface brightness of the QSO jet,
we can make a distinction between a point mass
(corresponding to an IMBH) and an 
extended structure (corresponding to a CDM subhalo).
It would be a unique probe of the IMCOs
whose nature has been under the veil of mistery.
\end{abstract}

\section{Introduction}

Gravitational lensing is a powerful tool for
constraining the amount of the dark matter
in the form of compact objects.
However, there is no stringent constraint on the 
intermediate mass compact objects (IMCOs) with a mass 
scale of  $10^{2-4}\ms$ and on the sub-lunar mass compact 
objects (SULCOs) with a mass scale 
of $<10^{-7}\ms$ (Inoue \& Tanaka 2003).

In particular, a method of directly detecting the 
IMCOs has been highly desired because the IMCOs are
important baryonic dark matter candidates as well. 
The IMCO can be an intermediate mass black hole 
or an intermediate mass cold dark matter (CDM) subhalo. 
The nature of both types
of IMCOs has not been understood well.

In fact, the abundance of these IMCOs can be much 
larger than expected. Recent observation 
of the ultra-luminous X-ray sources suggests a presence of 
IMBHs not only in the neighborhood of galaxy 
nucleus but also in the star clusters in the galactic halo far from 
the nucleus (Matsumoto et al 2001, Roberts et al 2004). 
Furthermore, the early reionization ($z\sim 20$) (Bennett et
al. 2003) suggested by the WMAP observation of the 
temperature-polarization
correlation may indicate strong UV radiation from massive first stars at 
the early period ($z \sim 20$) with a top-heavy IMF (Cen, 2003)
or that from a large number of micro-QSOs (Madau, et al. 2004),
which imply the existence of a large number of IMCOs.
The IMBHs are certainly important building blocks for
making the super massive black holes (SMBHs). Unfortunately,
their evolution has been under the veil of mistery.
Measurements of the abundance and the spatial distribution of the IMBHs 
inside the galactic halo would shed a new light on the evolution
history of the SMBHs. 

Similarly, we do not know anything about 
the abundance of the intermediate mass  $10^{2-5}\ms$ CDM subhalos.
Although they may suffer the tidal breaking owing to the gradient in the
gravitational potential of the galaxy halo, 
any compact objects whose size is smaller than 
the tidal radius can survive. Because a calculation 
of the survival probability during the major merger of galaxies 
is an intractable problem(see also Taylor \& Babul 2005), 
we need to observationally constrain the abundance of the CDM subhalos.

To do so, we propose to observe radio-loud QSO-galaxy strong 
lens systems with a submilli-arcsecond resolution, which will be achieved
 by the next space VLBI missions sush as the VSOP2
(Hirabayashi et al).  Then the submillilensing 
effects by IMCO perturbers can be directly 
measured (Inoue \& Chiba 2003).

\section{Theory}
The Einstein angular radius of a point mass with a mass 
$M$ is written in terms of 
the angular diameter distance to the lens $D_L$,
to the source $D_S$, between the lens and the source 
$D_{LS}$ as 
\BE
\theta_E\sim3\times10^{-2} \Biggl ( \f{M}{10^2 M_{\odot}}
\Biggr)^{\f{1}{2}} \Biggl ( \f{D_L D_S /D_{LS}}{\textrm{Gpc}}
\Biggr)^{-\f{1}{2}} \textrm{mas}.
\EE
Therefore, the radio interferometer with a resolution of 
0.04 mas can resolve the distortion of the image within 
the Einstein ring for a point mass $M \sim 200\ms$. 
As is well known, the surface density 
within the Einstein radius of the macrolens is nearly
critical. Therefore, the possibility of 
gravitational submillilensing by the IMCOs within
the Einstein ring of the macrolens ($=$galaxy halo) is 
very large. Furthermore, in the high frequency band
($>40$GHz), any absorption owing to the plasma along the
line of sight to the jets is suppressed.  
Thus, a high-resolution radio mapping 
of the QSO-galaxy lens systems is an 
ideallistic method for directly detecting the IMCOs. 
\section{Model}
To estimate the observational feasibility, 
we consider a simple model of a typical QSO-galaxy 
lensing system B1422+231, which consists 
of three magnified images A,B,C and a 
demagnified image D. The redshifts of the source and the lens
are $z_S=3.62$ and $z_L=0.34$, respectively. 
To model the macro-lensing, we adopt
a singular isothermal ellipsoid (SIE) in a constant
external shear field in which the
isopotential curves in the projected surface perpendicular to the
line of sight are ellipses (Kormann et al. 1994).
The perturbing IMCOs are modeled as point masses. 

\section{Simulation}
The surface number density of the perturbers 
near the Einstein ring for the macrolens 
with a radius ($r_E\sim 3.8$kpc) is 
found to be $N\sim f \times 2.4\times 10^1 
(M/10^3\ms)^{-1}$mas$^{-2}$ where $f$ denotes
the ratio of the surface mass density of 
the perturbers to that of the dark halo.
If we assume that the area of the magnified image of the 
QSO jet is $(10\textrm{pc})^2$, then we expect 
one IMBH in the line of sight to the jet of B1422+231 
if $f=0.003$. On the one hand, the total mass 
within the Einstein radius $r_E\sim 3.8$kpc 
is $\sim 10^{11}\ms$.
On the other hand, the total baryoic mass
within the Einstein radius of the macrolens is $\sim 10^{10}\ms$. From 
the Magorrian relation, the mass of the 
SMBH at the nucleus is expected to be $\sim 10^{7-8}\ms$.
Thus, we can find an IMBH in the line of sight 
to the jet if the total mass of the IMBH 
$\sim 0.003 \times 10^{11} \ms$ 
is comparable to the mass
of the SMBH provided that the IMBHs traces the 
mass distribution of the dark halo of the macrolens. 
In the extreme case where the IMBHs 
constitute the entire macrolens halo, 
the submillilensing effects owing to the IMBHs 
are very distinct (figure 1).
We would observe a several hundreds of monopole-like or
dipole-like distortion patterns in the surface brightness profile
of the QSO jet if observed with a submilli-arcsec resolution.
\begin{figure}[t]
\IG[width=17cm]{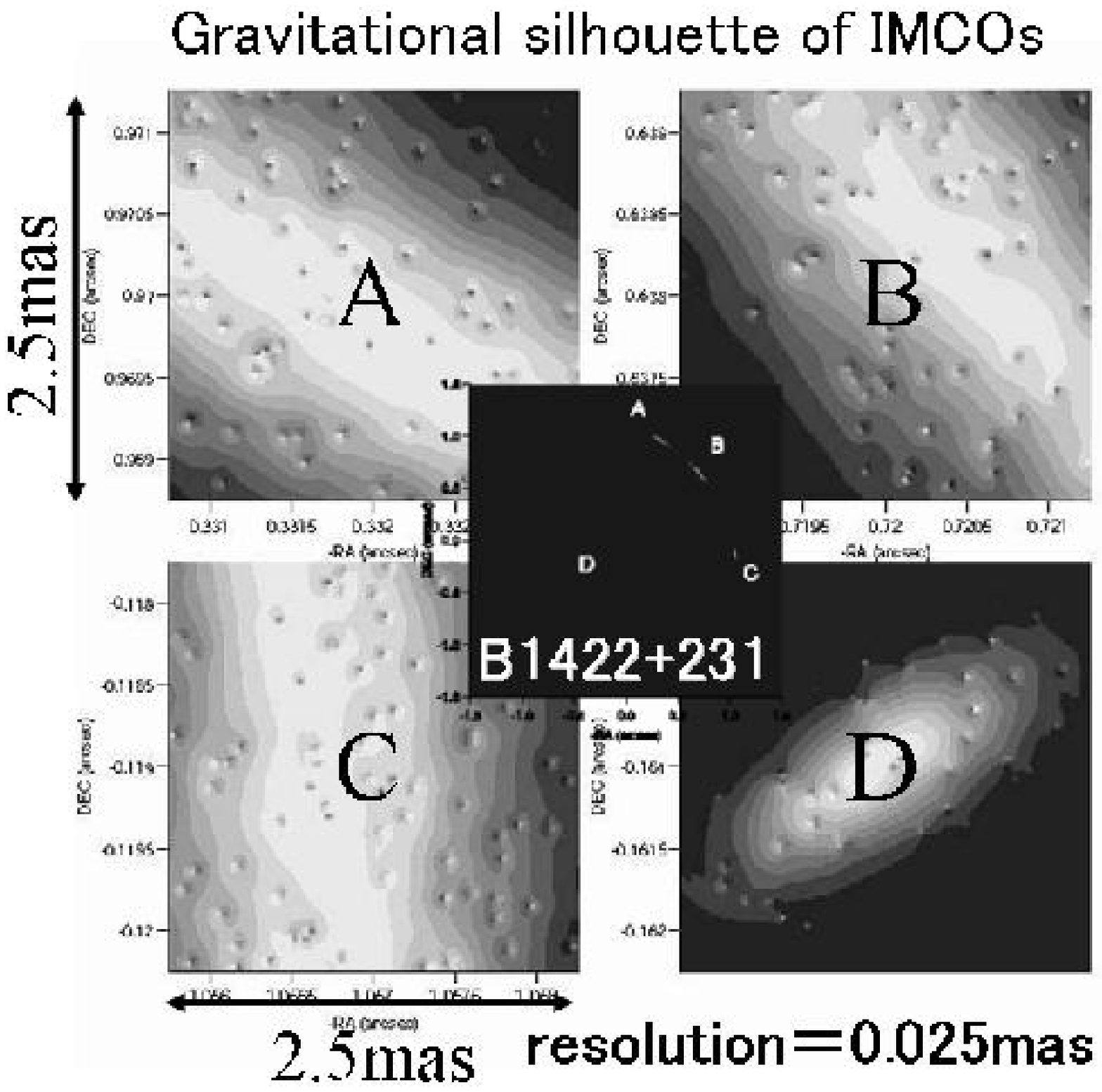}
\caption{Simulated zoomed-up multiply lensed images 
A,B,C and D in the B1422+231 (center) where 
the IMBHs constitute the entire macrolens CDM halo
$\Omega_{\textrm{CDM}}=\Omega_{\textrm{IMBH}}$.
  The silhouette of the 
IMBH would appear as a monopole-like 
or a dipole-like distortion pattern against the radio emission from the 
QSO. We assumed that the radius of the radio emitting region 
is $r=100$pc and the IMBH mass is set to $M=10^3\ms$. }

\end{figure}

\section{Extended-source effect}
If the the perturber is 
spatially extended (e.g. a singular isothermal sphere), then
the lensing effect is different from that of a point mass.
The density profile of the perturber can be reconstructed
from the local mapping between the observed image
and the non-perturbed image obtained from the 
prediction of the macrolens. In fact, the power of the 
radial density profile of the perturbers can be reconstructed from the 
perturbed image within the Einstein ring of the perturber
(Inoue \& Chiba 2005b). Then we can make a distinction
between the IMBH and the CDM subhalo. 
Furthermore, from the distortion outside the Einstein ring
of the perturber, the degeneracy 
between the perturber mass and the distance 
can be broken provided that the Einstein radius
of the perturber is sufficiently smaller than that 
of the macrolens(Inoue \& Chiba 2005a).
The precise measurement of the spatial variation in the surface
brightness of the QSO jet will provide us a 
large amount of information about the 
mass, abundance, and the spatial distribution of IMCOs.
\section{Summary}
If there are a sufficent number of IMCOs
inside the macrolens galactic halo, then 
we will be able to directly detect the 
silhouette of the IMCOs using the next 
space interferometers with a submilli-arcsec resolution 
in the radio band.
By measuring the local distortion of the QSO 
jet, we will be able to determine the density profile of the IMCOs.
The direct detection of the IMBHs or the CDM subhalos
will shed a new light on the 
formation process of the SMBHs and the reionization process  
which have been under the veil of mistery.

\end{document}